\documentclass[prb, aps, a4paper,twocolumn,showkeys,preprintnumbers]{revtex4-1}

\usepackage{graphicx}
\DeclareGraphicsExtensions{.eps, .png, .jpg, .pdf}
\graphicspath{{newfigures/}{Figures/PDF/}{Figures/}{./}}
\usepackage{epsfig}

\usepackage{longtable}
\usepackage{dcolumn} 
\usepackage{bm} 
\usepackage{amsmath} 
\usepackage{amssymb} 
\usepackage{subfigure}

\usepackage[ plainpages = false, pdfpagelabels,
                 pdfpagelayout = OneColumn, 
                 bookmarks,
                 bookmarksopen = true,
                 bookmarksnumbered = true,
                 linktocpage,
                 pagebackref,
                 colorlinks = true,
                 linkcolor = blue,
                 urlcolor  = blue,
                 citecolor = red,
                 anchorcolor = green,
                 hyperindex = true,
                 hyperfigures
                 ]{hyperref}

\begin{document}




\title{A theoretical study of the dynamics of atomic hydrogen on graphene bilayers}

\author{
\firstname{Mohammed}
\surname{Moaied}
}
\email{moaied5@yahoo.com}
\affiliation{Departamento de F\'{i}sica de la Materia Condensada, Universidad Aut\'{o}noma de Madrid, Cantoblanco, 28049 Madrid, Spain and Department of Physics, Faculty of Science, Zagazig University, 44519 Zagazig, Egypt.}

\author{
\firstname{J. A.}
\surname{Moreno}
}
\email{joseantonio.moreno@estudiante.uam.es}
\affiliation{
Departamento de F\'{i}sica de la Materia Condensada, Universidad Aut\'{o}noma de Madrid, Cantoblanco, 28049 Madrid, Spain.
}
\author{
\firstname{M. J.}
\surname{Caturla}
}
\email{mj.caturla@ua.es}
\affiliation{
Departamento de F\'{i}sica Aplicada, Universidad de Alicante, San Vicente del Raspeig, 03690 Alicante, Spain.
}

\author{
\firstname{F\'elix}
\surname{Yndur\'ain}
}
\author{
\firstname{J. J.}
\surname{Palacios}
}
\email{juanjose.palacios@uam.es}

\affiliation{
Departamento de F\'{i}sica de la Materia Condensada, Instituto Nicol\'as Cabrera (INC), and Condensed Matter Physics 
Institute (IFIMAC), Universidad Aut\'{o}noma de Madrid, Cantoblanco, 28049 Madrid, Spain.
}


\date{\today}


\begin{abstract}
We present a theoretical study of the dynamics of H atoms adsorbed on graphene bilayers with Bernal stacking. First, through extensive density functional theory calculations, including van der Waals interactions, we obtain the activation barriers involved in the desorption and migration processes of a single H atom.  These barriers, along with attempt rates and the energetics of H pairs, are used as input parameters in kinetic Monte Carlo simulations to study the time evolution of an initial random distribution of adsorbed H atoms. The simulations reveal that, at room temperature, H atoms occupy \textit{only one sublattice} before they completely desorb or form clusters. This sublattice selectivity in the distribution of H atoms may last for sufficiently long periods of time upon lowering the temperature down to $0^{\rm o}$C.  The final fate of the H atoms, namely, desorption or cluster formation, depends on the actual relative values of the activation barriers which can be tuned by doping. In some cases a sublattice selectivity can be obtained for periods of time experimentally relevant even at room temperature. This result shows the possibility for observation and applications of the ferromagnetic state associated with such distribution.
\end{abstract}



\maketitle


\section{Introduction}
Hydrogenation of carbon-based materials such as graphene, graphite, or carbon nanotubes is attracting much interest as a practical methodology to manipulate the electronic and magnetic properties of these materials in a reversible way. Hydrogenation of graphene, for instance, was found, both theoretically and experimentally, to be an effective way to turn this system from a gapless semiconductor into a gapful one with a tunable band gap\citep{D.C.Elias01302009,Sessi2009, Haberer2010, Yang2010, Balog2010,Lin15}. Controlled hydrogenation, on the other hand, has been predicted to induce interesting magnetic states with potential applications in spintronics \citep{Zhou2009,PhysRevB.81.165409,nn200558d,McCreary12,PhysRevLett.107.016602,Nair12}.
Unfortunately, the necessary  control has not been experimentally demonstrated to date.

Graphene is a single layer of carbon atoms bonded together in a bipartite honeycomb structure which is formed by two inter-penetrating triangular sublattices. The first neighbors of an atom in a given sublattice belong to the other sublattice and vice versa \citep{saito1998physical}. When atomic H is adsorbed on graphene the H atom bonds directly on top of a carbon atom and induces an intrinsic magnetic moment around the adsorption site with a 
net magnetic moment of 1 $\mu_B$\citep{PhysRevB.75.125408b,PhysRevB.77.035427, Casolo2009,PhysRevB.81.165409,Yndurain14}. Since the sublattices are chemically equivalent, the adsorption process is blind to the sublattice index. 
For graphite or multilayer graphene, unlike the monolayer case, surface carbon atoms in one sublattice present others underneath while the ones in the complementary sublattice  do not (assuming Bernal stacking). Thus, at first sight, it should not come as a surprise that the two sublattices offer different binding energies to H atoms, thus favouring adsorption on one of them (at least at low enough coverages).

The specific details of the
desorption and diffusion processes of the adsorbed H atoms are, nevertheless, essential to determine the final or temporary hydrogenation patterns and related electronic properties and, most importantly, the time scale for reaching such patterns. When H atoms are initially deposited, e.g., by cracking molecular H$_2$\cite{Güttler04}, it is expected that they will reach both sublattices with equal probability. The electronic state thus induced will correspond to that of a non-magnetic insulator for large concentrations\citep{D.C.Elias01302009,Sessi2009, Haberer2010, Yang2010, Balog2010}, an antiferromagnet for intermediate concentrations\cite{PhysRevB.77.195428}, or a paramagnet for low concentrations\cite{Nair12}. At room temperature, which for practical applications is the most interesting case, both desorption and diffusion processes are, in principle, active\citep{Jeloaica99, PhysRevLett.97.186102,Chen2007,Denis2009,Kerwin2008, Ivanovskaya2010,Sha2002,Herrero09,Huang11,Huang11-1,Borodin11}. If desorption rates are larger than diffusion ones, the sample will loose H from both sublattices, but at different rates. If, on the other hand, diffusion or migration rates are larger than desorption ones, H atoms will move across the surface performing many jumps before desorbing. In this case they will certainly spend more time on one sublattice than on the other. In both scenarios one sublattice may become more populated than the other, although in the second one the H atoms will also have more chances to come across one another and form stable non-magnetic clusters\citep{Ferro03,Casolo2009,Sljivancanin2009,
Roman07,Andree06}.

Here we show that one can get, at least temporarily, a nearly 100\%  sublattice selectivity in the adsorption, i.e., a distribution where all adsorbed H atoms occupy the same sublattice on the surface graphene monolayer. At room temperature this may occur for periods of time of minutes. Upon lowering the temperature down to $0^{\rm o}$C, the single-sublattice distribution may, however, survive for days. This result does not qualitatively depend on the specific values of the migration and desorption barriers. For instance, upon changing the carrier concentration of the bilayer system, which, in turn, changes the magnitude of these barriers\cite{Huang11-1}, we always obtain such temporary distribution of H, only the final fate of the atoms  being affected. For hole doping, all atoms eventually form dimers or clusters while for electron doping (or no doping) all atoms eventually desorb. This remarkable result has important implications since theory predicts that when all the H atoms bind to the same sublattice the resulting electronic ground state should be a ferromagnetic state\citep{PhysRevLett.62.1201,Zhou2009,PhysRevB.77.195428} with a typically very high Curie temperature for a wide range of concentrations\cite{Moaied14}. 

\section{Computational Methodology}

We combine two types of computational methodologies. On the one hand, density functional theory (DFT)\citep{PhysRev.136.B864, PhysRev.140.A1133} and, on the other, kinetic Monte Carlo (KMC) simulations. The essential ingredients for the latter, i.e., activation energies and attempt rates, are obtained from the former. A graphene bilayer is considered all along; the results thus obtained need not be exactly representative of the physics of multilayer graphene or graphite, although arguments can be put forward to carry our results over to these cases.

\subsection{van der Waals DFT with SIESTA and preliminary checks}
As we are dealing with weakly interacting graphene layers where dispersion (van der Waals) forces due to long-range electron correlation effects play a key role, we employ the non-local van der Waals density functional (vdW-DF) of Dion et al. \citep{PhysRevLett.92.246401} as implemented by Rom\'{a}n-P\'{e}rez and Soler\citep{PhysRevLett.103.096102, PhysRevLett.103.096103} in the SIESTA code\citep{PhysRevB.53.R10441, 0953-8984-14-11-302}. To describe the interaction between the valence and core electrons we used norm-conserved Troullier-Martins pseudopotentials \citep{PhysRevB.43.1993}. To expand the wave functions of the valence electrons, a double-$ \zeta $ plus polarization (DZP) basis set was used \citep{PhysRevB.64.235111}. We experimented with a variety of LCAO basis sets and found that including polarization basis elements was important; however, a triple-$ \zeta $ (TZP) basis set produced results essentially indistinguishable from those obtained with DZP.  For the Brillouin zone sampling we used at least a $ 20\times 20 $ Monkhorst-Pack $k-$mesh, increasing the density of $k$ points for occasional checks.  We ensured that the vacuum space is at least 25 \AA{} so that the interaction between functionalized layers and their periodic images can be safely ignored. We have also checked that the results are well converged with respect to the real space grid. The atoms are allowed to relax down to a force tolerance of 0.020 eV/\AA, keeping the necessary coordinates of the H atom fixed to obtain the corresponding desorption energy curves and migration landscapes. Spin polarization was included in the calculations because hydrogenation induces magnetism in single-layer as well as multilayer graphene\cite{Moaied14}.

\begin{figure}[!htbp]
    \centering{\includegraphics[width=3.5in]{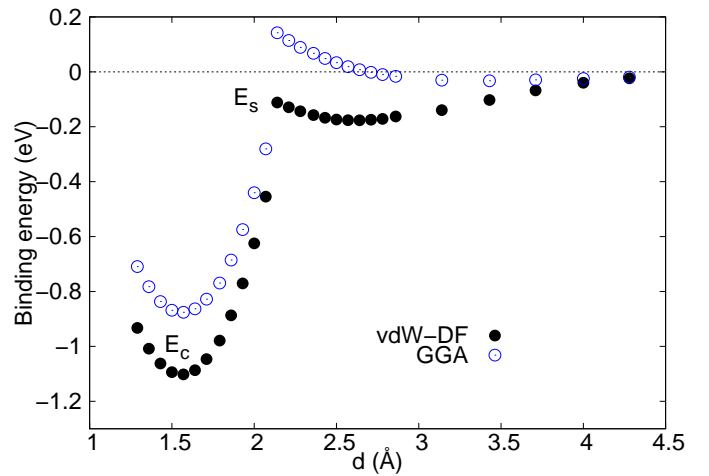}}
    \caption{(color online). Binding energy of a H  atom on top a C atom in monolayer graphene (a $4\times4$ supercell) as a function of the distance to the graphene plane. Blue dots correspond to a standard GGA functional while black ones correspond to the vdW-DF, as explained in the text.}
    \label{binding-monolayer}
\end{figure}

Before addressing the specifics of bilayer graphene in Sec. \ref{bilayers}, we will briefly revisit the monolayer graphene case. H atoms are known to preferentially adsorb on top of carbon atoms. Figure \ref{binding-monolayer} shows the vdW-DF binding energy (black dots) of a H atom on top a C atom for a \textit{single} graphene layer as a function of the distance to the graphene plane, $d$. The binding energy is defined as usual:
\begin{equation}
E_{\rm bind}(d)=E_{\rm graphene+H}(d) - E_{\rm H} - E_{\rm graphene}.
\end{equation} 
(As a complementary accuracy check, we have made sure that the energy of the isolated H atom, $E_{\rm H}$, is 1 Ryd.) Two distinctive minima or adsorption states appear: a strongly bound chemisorption state at  $\approx 1.6$ \AA \hspace{0.05cm} with binding energy $E_{\rm c}$ and a weakly bound physisorption state, which can be appreciated as a shallow minimum around $\approx 2.6 $\AA. The distance between the host C atom and the H atom abruptly changes in between both minima, being $\approx 1.3$ \AA \hspace{0.05cm} at the chemisorption state and $\approx 2.2$ \AA \hspace{0.05cm} at the physisorption state.  The derivative discontinuity at the transition point between energy minima can be attributed to the mean field treatment which could be smoothed out by more sophisticated methods which do not break spin symmetry\cite{Casolo2009}. 

The importance of using a vdW-DF not only reflects on the fact that standard GGA functionals do not bind (or barely bind) graphene layers into a bilayer or graphite. Blue dots in Fig. \ref{binding-monolayer} correspond to the same calculation using a commonly used GGA functional\cite{Zhang98}. The result is essentially similar to many others found in the literature\citep{Jeloaica99, PhysRevLett.97.186102,Chen2007,Denis2009,Kerwin2008, Ivanovskaya2010,Sha2002,Herrero09,Huang11,Huang11-1,Borodin11}, whereas it is quantitatively and even qualitatively different from the vdW-DF result. In particular, deeper chemisorption and  physisorption minima are obtained in the vdW-DF calculation. Although the saddle point separating  the chemisorption state from the physisorption one is not smooth in our numerics, we can still appreciate that it has a negative value $E_{\rm s}\approx -150$meV. Most calculations in the literature  exhibit slightly positive saddle point energies\citep{Jeloaica99, PhysRevLett.97.186102,Chen2007,Denis2009,Kerwin2008, Ivanovskaya2010,Sha2002}. Interestingly, the difference $E_{\rm s} - E_{\rm c}$ is similar for both functionals.

\subsection{Object kinetic Monte Carlo algorithm}
The KMC calculations have been performed to understand the time evolution of the surface distribution of the H atoms for different temperatures and  concentrations. KMC algorithms are powerful techniques to study the dynamics of a system of particles when the different events that those particles can perform are known as well as their probabilities (for a recent review see \citep{Voter2007}). There are many different algorithms with the name KMC. In this case we use what is often known as an object kinetic Monte Carlo (OKMC) algorithm, based on the residence time algorithm or Bortz-Kalos-Liebowitz (BKL) algorithm \citep{Kalos:1986:MCM:7050,Voter2007}  Briefly, in an OKMC algorithm a list of possible events is defined with a given probability for each event, $\Gamma_{i}$. This probability usually follows an Arrhenius dependence with temperature:
\begin{equation}
 \Gamma_{i} = \Gamma_{i}^{0} exp {\left(\frac{-\Delta_{i}}{k_{B} T}\right)},
\label{Arrhenius}
\end{equation}
where $k_{B}$ is the Boltzmann constant, $\Delta_{i}$ is the activation energy of the given event,
 and $\Gamma_{i}^{0}$ is the attempt frequency. In this case the activation energies are related to migration or diffusion energies, desorption, and dissociation energies. 

The total rate for all events, $R$, is then calculated as:
\begin{equation}
 R = \displaystyle\sum\limits_{i=1}^{n_{\rm e}} \Gamma_{i} N_{i},
\end{equation}
where $n_{\rm e}$ is the total number of events and $N_{i}$ is the number of particles that can perform event $i$. An event is then selected by picking a random number between $[0, R]$. In this way one event is selected every Monte Carlo step from all possible with the appropriate weight.  Once the event has been selected, a random particle is chosen from all those that can undergo that event. The particle is then moved and the total rate has to be computed again for the next simulation step. At every Monte Carlo step the time increases by:
\begin{equation}
 t = \frac{-log{\xi}}{R},
\end{equation}
where $\xi$ is a random number between $[0,1]$ that is used to give a Poisson distribution of the time. 

\begin{figure}[!htbp]
  \begin{center}
      \includegraphics[height=2.50in]{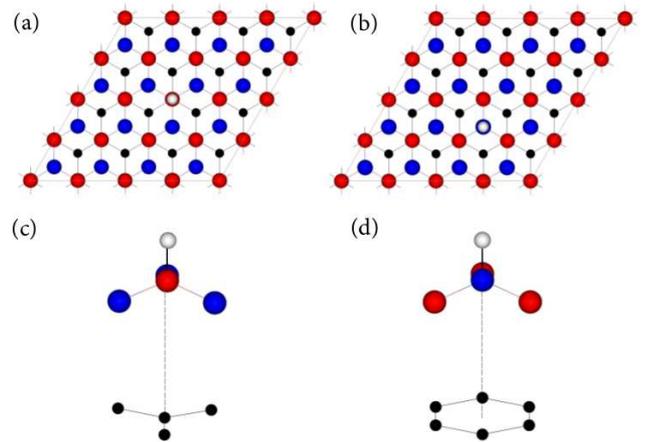}
    \caption{(color online). Atomic structure of H on bilayer graphene. $\alpha$ (a) and $\beta$ (b) sites top view and $\alpha$ (c) and $\beta$ (d) sites side view detail.}
    \label{H-over-bilayer}
  \end{center}
\end{figure}

\section{Activation barriers for a H atom on bilayer graphene}
\label{bilayers}
\subsection{Desorption barriers}
\begin{figure}
    \centering
    {
        \includegraphics[width=3.5in]{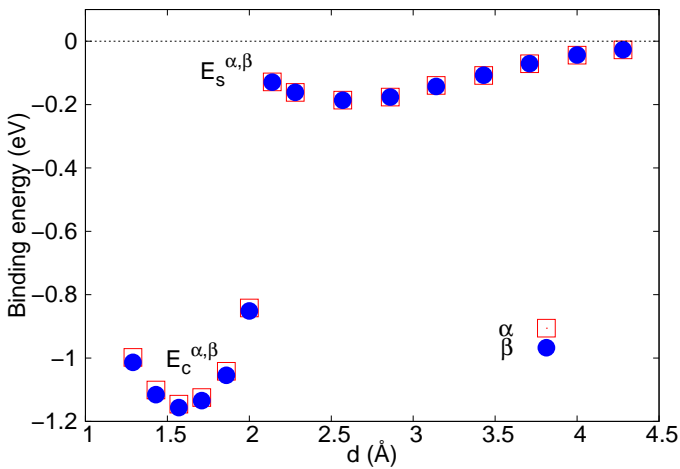}        
    }
    \caption{(color online). Binding energy curves for a H  atom  adsorbed on the $\alpha$ (red) and $\beta$ (blue) sites of a bilayer graphene surface (a $4\times4$ supercell).}
    \label{binding-bilayer}
\end{figure}

The physics of atomic H adsorption on a bilayer is not much different from that on monolayer graphene. Now only the upper layer is allowed to relax while the C atoms in the lower layer are being fixed at their equilibrium positions, simulating the presence of a substrate such as  graphite or SiC. Due to the chosen Bernal stacking, the two sublattices are not equivalent any more. To stress this important point, we will denote the two different adsorption sites as $\alpha$ and $\beta$ from now on (see Fig. \ref{H-over-bilayer}).  The vdW-DF binding energy curves corresponding to both adsorption sites are presented in Fig. \ref{binding-bilayer} for a $4\times4$ supercell. As for the monolayer case (see Fig. \ref{binding-monolayer}), both curves exhibit two minima or adsorption states. A magnetic moment of 1$\mu_{\rm B}$ also appears on the surface layer at the chemisorption state while it transfers to the H atom at the physisorption minimum. In Fig. \ref{H-over-bilayer} we show the  atomic structure for the chemisorption state.  The characteristic $sp^3$ re-hybridization induced by the H atom is patent in both adsorption sites. The resulting atomic structures are not identical (although this can barely be appreciated in the figure) and the chemisorption energies, $E_{\rm c}^{\alpha}$ and $E_{\rm c}^{\beta}$,  are slightly different as well ($|E_{\rm c}^{\alpha}| < |E_{\rm c}^{\beta}|$). No significant difference between adsorption sites can be appreciated in the physisorption part of the curves (see Fig. \ref{binding-bilayer}). 

\begin{figure}[!htbp]
    \centering{\includegraphics[width=3.5in]{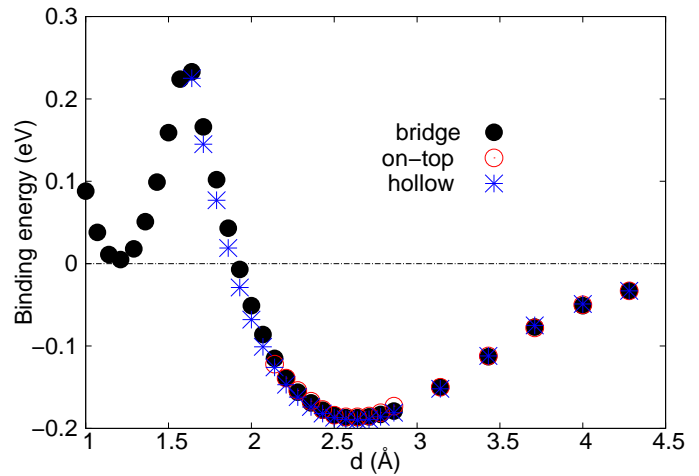}}
    \caption{(color online). Detail of the physisorption binding energy curve for a H  atom on top a C atom ($\beta$ site), on hollow, and on bridge position for a bilayer graphene (a $4\times4$ supercell).}
    \label{physisorption}
\end{figure}

Similarly to the monolayer case, the saddle points separating chemisorption from physisorption minima, $E_{\rm s}^{\alpha}$ ($\approx E_{\rm s}^{\beta} $), present negative values ($\approx -150 $ meV).  In the light of this result one might wonder whether the desorption activation barriers, $\Delta E^{\alpha(\beta)}_d $, to be considered in the OKMC calculations, should correspond to $|E_{\rm c}^{\alpha(\beta)}|$ or to the smaller difference $E_{\rm s}^{\alpha(\beta)}- E_{\rm c}^{\alpha(\beta)}$. Figure \ref{physisorption} shows a blow-up of the physisorption minima for the H atom being on top a $\beta$ site, hollow, and bridge positions. As can be appreciated, the physisorption energy barely changes with the position of the H atom with respect to the substrate; in other words, the physisorption energy landscape is essentially flat on the scale of eV.  H atoms being thermally promoted from the chemisorption  to the physisorption state across the saddle point may freely wander across the surface before they are chemisorbed again or finally desorb. In what follows we will assume that the distance travelled  on the physisorption channel is long enough so that,  before the chemisorption process occurs again, the H atoms meet other H atoms or hit grain boundaries where they permanently stick (see below). Thus, to any practical purpose, we will consider  $\Delta E^{\alpha(\beta)}_d= E_{\rm s}^{\alpha(\beta)}- E_{\rm c}^{\alpha(\beta)}$. (The possibility of considering $|E_{\rm c}^{\alpha(\beta)}|$ as the relevant desorption activation barrier is briefly discussed below).

\subsection{Migration barriers}
\begin{figure}
    \centering
    \subfigure
    {
        \includegraphics[width=3.30in]{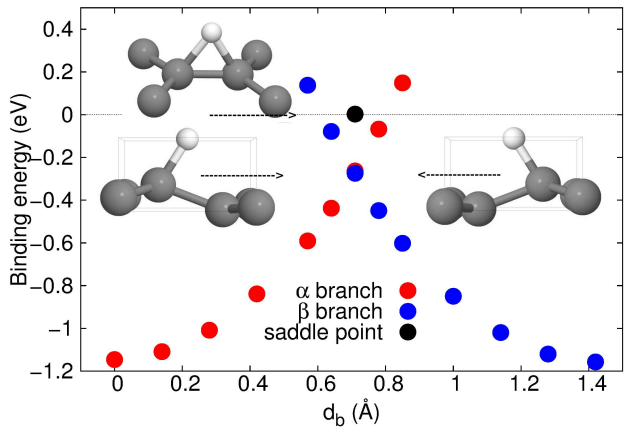}
    }
    \caption{(color online). Binding energy of a H atom on bilayer graphene placed along the bond that links the $\alpha$ site to the $\beta$ site for a 4x4 supercell. Despite the appearances, the crossing point is not a saddle point (see lower insets and see text). The true saddle point is depicted with a black dot and the associated structure is shown in the upper inset. (In all insets the bottom layer is not shown for clarity).}
    \label{migration}
\end{figure}

Figure \ref{migration} shows the binding energy obtained by displacing the H atom along the bond line ($d_b$) joining the $\alpha$ and $\beta$ sites (for a $4\times4$ supercell). These results are obtained by fixing the $x-y$  coordinates (plane) of the H atom, letting the $z$ coordinate (height) and all the positions of the carbon atoms of the top graphene layer to relax. It is important to notice that the sets of points originating from both sites (differentiated by colors) do not cross in the coordinates phase space. At the ``crossing'' point around $d_b=0.7$ \AA, two different solutions with very different atomic structures are obtained, one where the H atom remains bonded to the $\alpha$ C atom and the other one where the H atom is bonded to the $\beta$ C atom (see insets). In fact, as shown in the figure, both sets of points can be smoothly continued beyond the crossing point. While the actual path in coordinates phase space for the H atom to move from the $\alpha$ site to the $\beta$ one (or viceversa) is not known to us, it must cross a saddle point where the distance between the H atom and the two C atoms involved is the same. The binding energy of this saddle point is $\approx 0.0$ meV (represented by a black dot in Fig. \ref{migration}) and its atomic structure is shown in the upper inset. Taken with respect to the $\alpha$ and $\beta$ chemisorption minima, the saddle point energy gives the activation barriers to directly migrate between sites, $\Delta E_{\rm m}^{\alpha\leftrightarrow\beta}$. From the figure we see that these barriers are essentially equal to the respective chemisorption energies $|E_{\rm c}^{\alpha(\beta)}|$, but are larger than the desorption barriers as defined above (by $\approx 150$ meV). Importantly, as shown in Fig. \ref{physisorption}, the H atom is still considerably bound to the surface at the bridge or saddle point position, being thus possible to directly migrate between sites without moving to the physisorption channel or desorbing at room temperature.

\subsection{Activation processes involving two or more H atoms}

The activation energies discussed above dramatically change when the H atom is close to other H atoms. 
Here we will make use of the energetics of two H atoms on a graphene monolayer 
(as reported, e.g., in Ref. \onlinecite{Bachellerie07,Rougeau06,Sljivancanin2009,Moaied14}) to estimate how the 
vicinity of other atoms modify these barriers.  A thorough study for some cluster possibilities can be found in Ref. \onlinecite{Sljivancanin2009} which also serves of guidance for the following considerations. 
Our basic assumption is that when a H atom attempts to break an $\alpha-\beta$ ``bond" 
[two H atoms sitting on nearest-neighbour C atoms, also known as orto (O) dimer\cite{Sljivancanin2009}] 
an extra pair binding energy $E_b \approx 1.4$ eV  
has to be paid\cite{Moaied14} in addition to the migration barrier calculated for isolated atoms. 
In other words, the energy required to change an $\alpha-\beta$  H dimer into an $\alpha-\alpha$ or $\beta-\beta$  dimer [metastable (M) dimers\cite{Sljivancanin2009}] is given by
\begin{equation}
\begin{split}
\Delta E^{\alpha\leftrightarrow\beta}_{m'} &= \Delta E^{\alpha\leftrightarrow\beta}_{m}+ E_b \\ 
\end{split}
\label{clusterbarriers}
\end{equation}
The same assumption will be made for migration processes involving the breaking of $\alpha--\beta$ pairs [also known as para (P) dimers (see Ref. \onlinecite{Sljivancanin2009})], which are also strongly bound with a binding energy of $\approx 1.35$ eV\cite{Moaied14}. 
Finally, we also apply the same addition rule to the desorption barrier of a H atom if, 
in the desorption process, an $\alpha-\beta$ or $\alpha--\beta$ bond is broken:
\begin{equation}
\Delta E^{\alpha(\beta)}_{d'}= \Delta E^{\alpha(\beta)}_{d}+ E_b.
\end{equation}

In order not to complicate in excess and unnecessarily the OKMC simulations, additional assumptions regarding the energetics need to be made: a) We will ignore the binding energy or attraction between H atoms exerted at distances longer than those in the above referred pairs; b) H clusters (more than two atoms in close proximity) can always be considered as composed of dimers linked by $\alpha-\beta$ and/or $\alpha--\beta$ bonds so that when a H atom attempts a migration or desorption, the activation barrier will be that of breaking the corresponding bond(s), regardless of the cluster structure and number of H atoms forming it; c) finally, we will assume that the attempt rates $\Gamma_i^0$ are not modified by the presence of other H atoms. The accuracy of all these estimates is in fact not critical at all to the final results. Once clusters are formed, the activation energies are so large that they never break apart (in a relevant time scale and for relevant temperatures). When interpreting the results, one should only keep in mind that ignoring the interaction between H atoms at longer distances might reduce the likelihood for formation of clusters.

\begin{table}[h]
\centering
\caption{Events included in the OKMC calculation along with the correspoding activation barriers (in eV) and attempt frequencies (in s$^{-1}$) for each type of event, as obtained from the DFT calculations. The factor in front of the frequency values relates to the number of available positions to jump to.}
 \begin{tabular}{c c c}
                \hline
               Event & barrier & frequency \\
                \hline
                  Isolated atom events & & \\
                \hline
                Migration from $\alpha$ site & 1.15 & 3$\times$ 3.5 $10^{13}$ \\
                Migration from $\beta$ site & 1.23 & 3$\times$ 3.5$ 10^{13}$ \\
                Desorption of $\alpha$ site & 1.00 & 7.10 $10^{13}$ \\
                Desorption of $\beta$ site & 1.08 & 7.10 $10^{13}$ \\
                \hline
                 Dissociative events & & \\
                \hline
                Migration from $\alpha$ in $\alpha-\beta$ dimer & 2.55 & 2$\times$ 3.5 $10^{13}$ \\
                Migration from $\beta$ in $\alpha-\beta$ dimer & 
2.63 & 2$\times$ 3.5 $10^{13}$ \\
                Migration from $\alpha$ in $\alpha--\beta$ dimer & 2.5 & 3$\times$ 3.5 $10^{13}$ \\
                Migration from $\beta$ in $\alpha--\beta$ dimer & 2.58 & 3$\times$ 3.5 $10^{13}$ \\
                Desorption of $\alpha$ in $\alpha-\beta$ dimer & 2.40 &  7.10 $10^{13}$ \\
                Desorption of $\beta$ in $\alpha-\beta$ dimer & 2.48 &  7.11 $10^{13}$ \\
                Desorption of $\alpha$ in $\alpha--\beta$ dimer & 2.35 &  7.10 $10^{13}$ \\
                Desorption of $\beta$ in $\alpha--\beta$ dimer & 2.43 &  7.11 $10^{13}$ \\
\hline
                \end{tabular}
                \label{TableKMC}
\end{table}

\section{Kinetic Monte Carlo simulations}

From simple statistical considerations, the fact that $\alpha$ and $\beta$ chemisorption energies are different and that both migration and desorption activation barriers are in the vicinity of $\approx 1$ eV anticipate that near room temperature H atoms may occupy $\alpha$ and $\beta$ sites with a significantly different probability. The question that actually needs to be addressed is how the two activated migration and desorption processes compete to determine the evolution of an initial random arrangement of H atoms and whether or not, in the end, all H atoms desorb or group together forming non-magnetic clusters, which, according to Eq. \ref{clusterbarriers} should be thermodynamically stable\cite{Šljivančanin201270}. 

These questions can be answered through OKMC calculations. We have thus implemented an OKMC algorithm that includes a total of 12 events, as shown in Table \ref{TableKMC}.   The rates for migration and desorption are estimated from the DFT activation
energies, as explained above, and from the attempt frequencies, calculated using simple harmonic models 
derived from the energy curves near the minima. 
Despite of our efforts to obtain accurate activation energies, it is unrealistic to expect a precision down to meV or even tens of meV, so the overall time scales may still be somewhat uncertain. However, the difference between the $\alpha$ and $\beta$ desorption and migration barriers is undeniable. This difference has been set to $  80$ meV, which approximately corresponds to the chemisorption energy difference in the zero concentration limit\cite{Moaied14}. The results are thus expected to be generically representative of low concentrations. In all the simulations we have considered a number of lattice sites in the range of $40000-80000$  in order to have access to low values of the coverages $C$ (defined as the ratio of H atoms to C atoms) for a sufficiently large number of H atoms ($100-500$). This keeps the statistical noise in the curves sufficiently low. 
 
The actual chemisorption process after cracking molecular H is unknown to us and probably needs a separate discussion. In what follows we will assume an initial random distribution of chemisorbed H atoms for the chosen concentration.  A H atom can either jump to a neighbouring location or desorb from the surface layer. After every jump it is necessary to check whether or not other H atoms are located in the vicinity. As explained above, we consider the formation of two types of pairs or dimers and clusters formed out of them which are kept immobile as a whole in the calculation. As a result we will show the time evolution of the relative abundance of H atoms adsorbed on each sublattice [$\alpha$ (in red) and $\beta$ (in blue)]. Solid lines will correspond to isolated atoms while H atoms belonging to $\alpha-\beta$ dimers will be represented by dashed lines and those associated with $\alpha--\beta$ dimers by dotted lines.  In fact, since clusters with more than two H atoms can be formed, we should generally speak about H atoms being part of bonds instead of dimers. For instance, a trimer such as $\alpha-\beta-\alpha$ contains two $\alpha$ atoms and one $\beta$ atom which we associate with $\alpha-\beta$ bonds  in our analysis. Therefore, as can be seen in some cases below, the number of $\alpha$ and $\beta$ atoms associated with these bonds does not need to be identical. Likewise, the count of H atoms associated with $\alpha--\beta$ dimers must be interpreted similarly. Cases such as $\beta$ atoms in clusters containing different types of bonds, e.g., $\alpha-\beta--\alpha$ are associated with both, but these seldom appear and do not significantly affect the count. We will also show  the number of desorbed atoms (black lines).

\begin{figure}
    \centering
    \subfigure{       
       \includegraphics[width=3.3in]{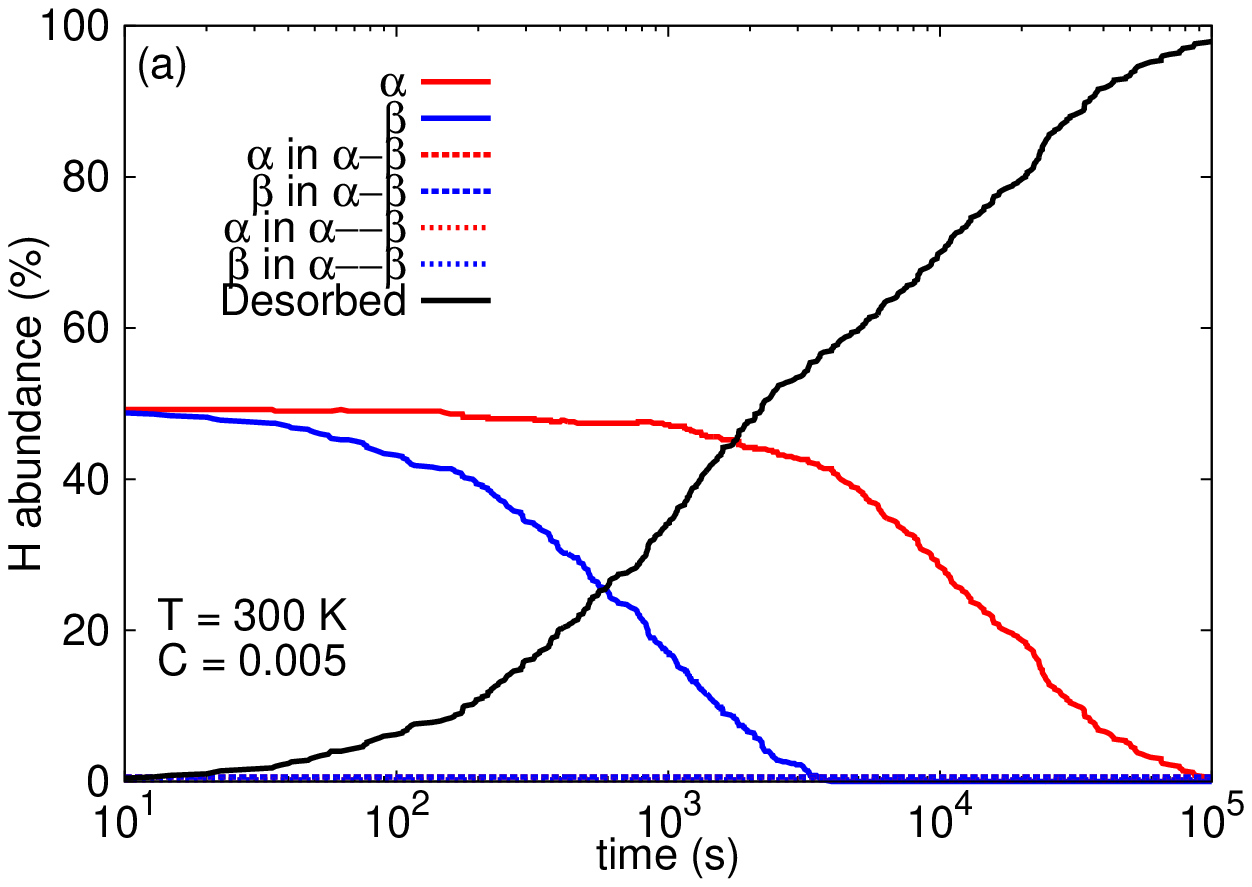}}
    \subfigure{
       \includegraphics[width=3.3in]{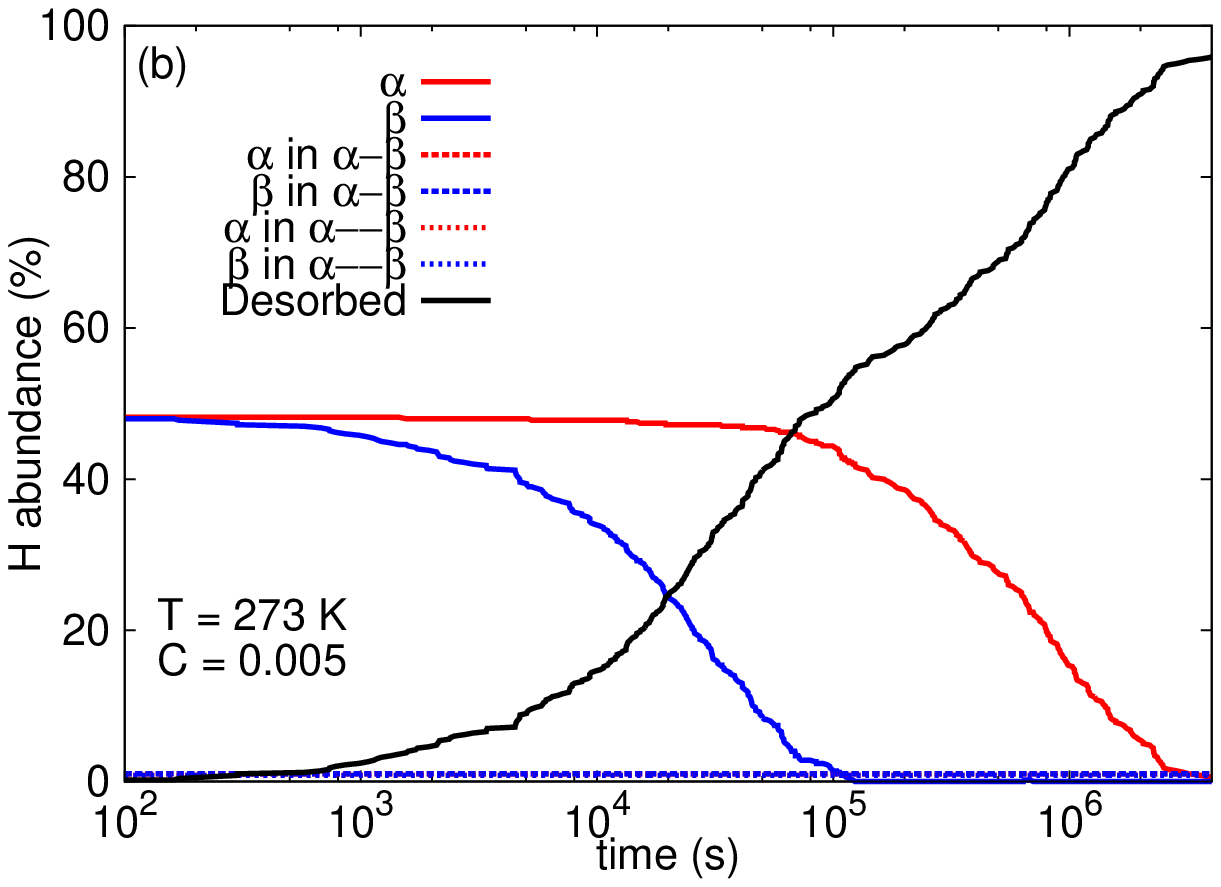}}
    \caption{(color online). Time evolution of the relative abundance of H atoms adsorbed on the two graphene sublattices for
    an initial concentration $C=0.005$ and  two representative temperatures (a) T=300 K and (b) T=273 K. Solid lines correspond to isolated H atoms while dashed and dotted lines correspond to H atoms forming part of dimers (or clusters) through short or long bonds, respectively (see text for a detailed explanation of these two types of bonds). Black solid line refers to desorbed atoms.}
     \label{bilayer-abundance}
\end{figure}

\begin{figure}
    \centering
    \subfigure{ 
       \includegraphics[width=3.3in]{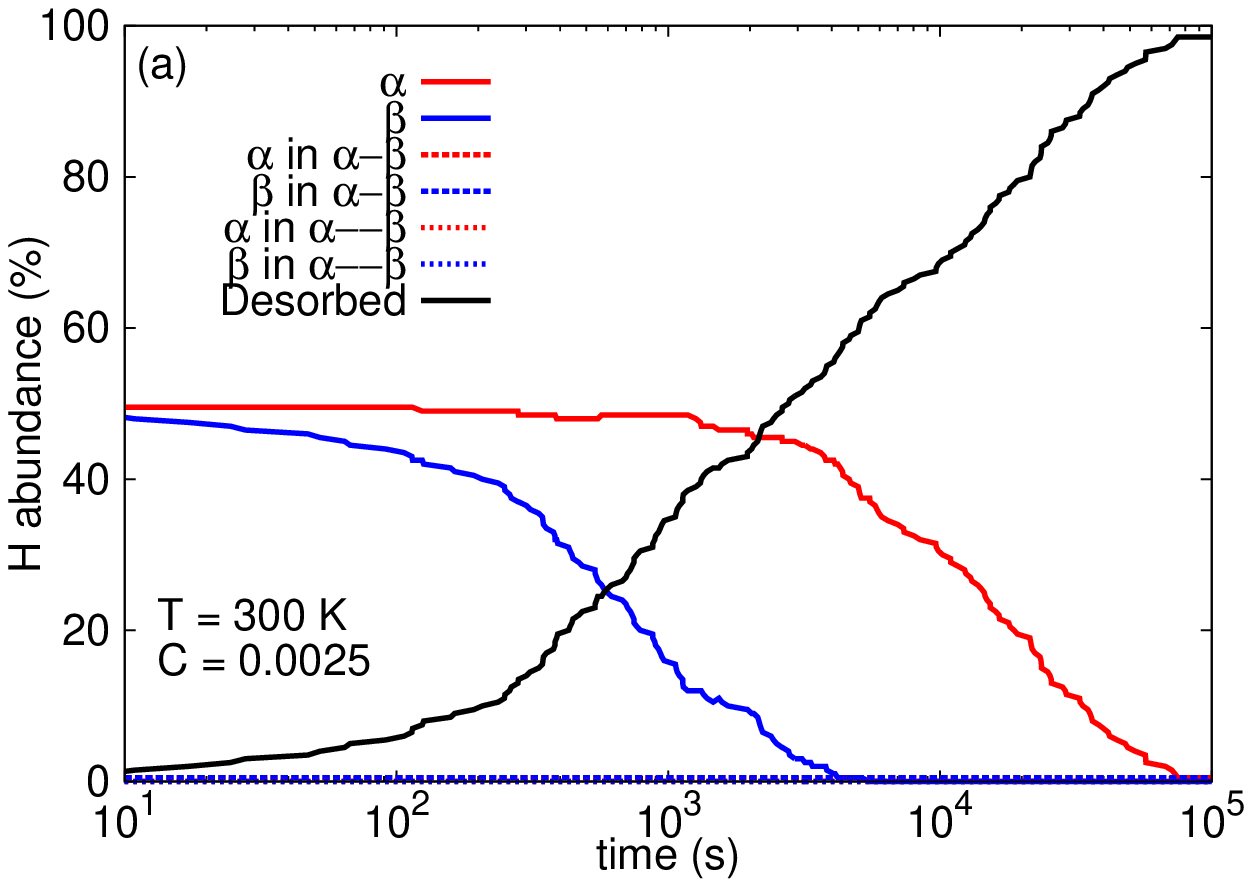}} 
       \subfigure{ 
       \includegraphics[width=3.3in]{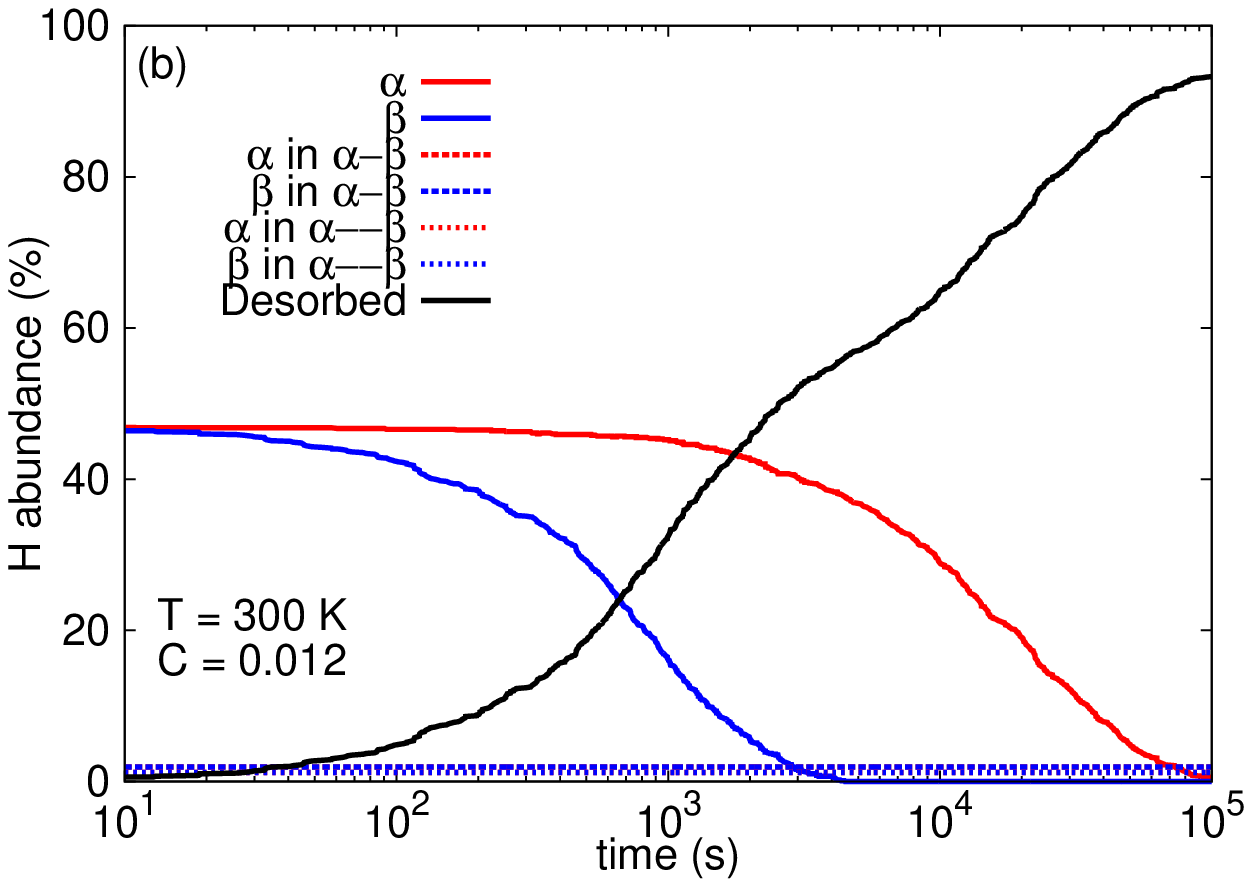}} 
    \caption{(color online). Same as in Fig. \ref{bilayer-abundance}, but only at room temperature for (a) $C=0.0025$ and (b) $C=0.0125$.}
     \label{bilayer2-abundance}
\end{figure}

\subsection{Neutral graphene bilayer}

Representative results for different initial values of the H concentration $C$ and different temperatures for a neutral bilayer are presented in  Figs. \ref{bilayer-abundance} and \ref{bilayer2-abundance}. In all cases we generically observe that the population of isolated H atoms on both sublattices decreases overtime due to desorption. Expectedly, after a certain time, 100\% of the remaining H atoms sit on the $\beta$ sublattice due to the larger desorption barrier of this site. At room temperature it takes $\approx$ 1 hour for this to occur  [see Fig. \ref{bilayer-abundance}(a)] and after a few hours most of H has desorbed.  At lower temperatures ($T=273$ K) the time window to have a full concentration of H atoms on the same sublattice logically increases [to approximately several days as shown in Fig. \ref{bilayer2-abundance}(b)]. Interestingly, this single-sublattice distribution now lasts for hours before desorption takes over. In both cases cluster formation barely occurs because of the significant difference in the desorption and migration activation barriers. (However, we should keep in mind that we have excluded the possibility of migration on the physisorption channel which might increase the probability of cluster formation.) The whole picture remains essentially the same for a wide range of H concentrations, as shown in Fig. \ref{bilayer2-abundance}. Apart from an obvious statistical smoothing of the curves for larger concentrations,  the probability for an initial accidental  formation of dimers or clusters is, as expected, larger for larger $C$, although the number of clusters does not change overtime.

\begin{figure}
    \centering
        \includegraphics[width=3.3in]{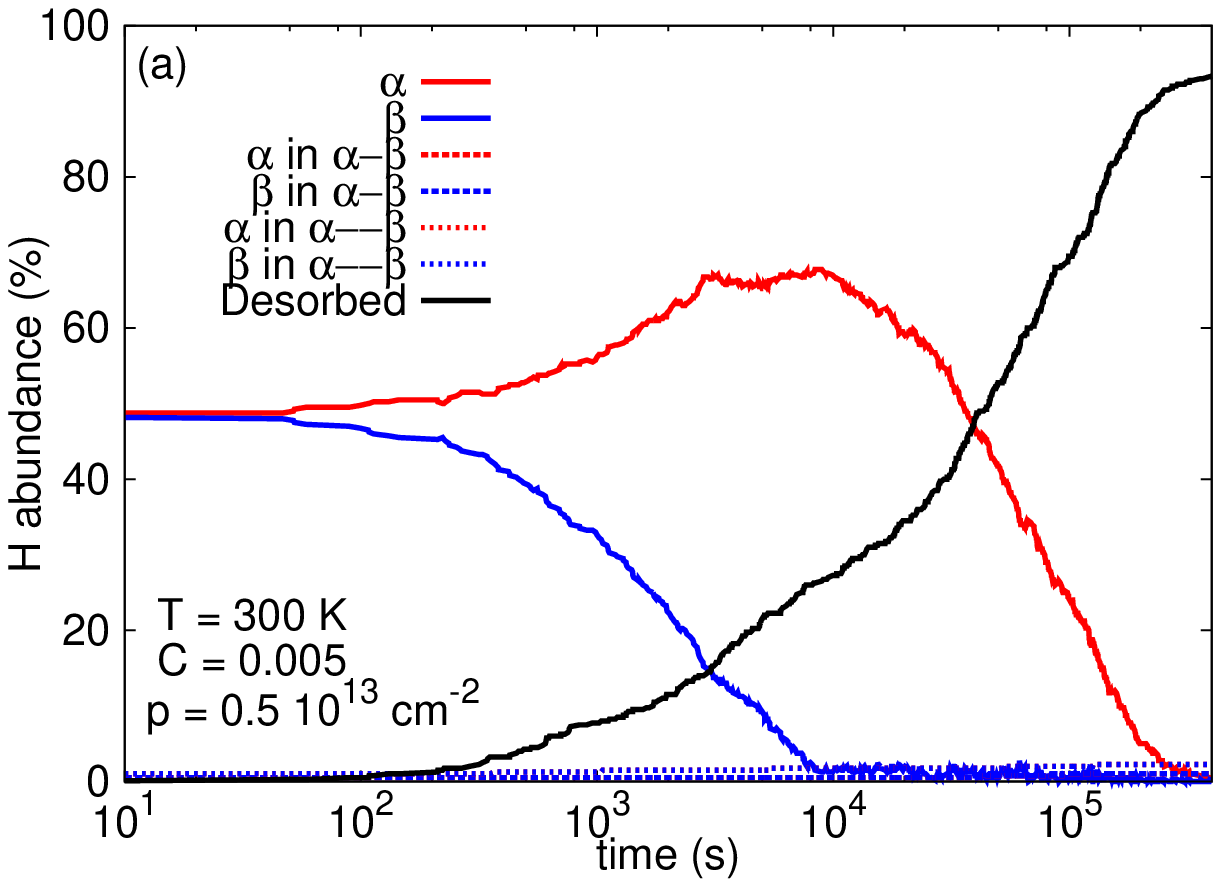}
        \includegraphics[width=3.3in]{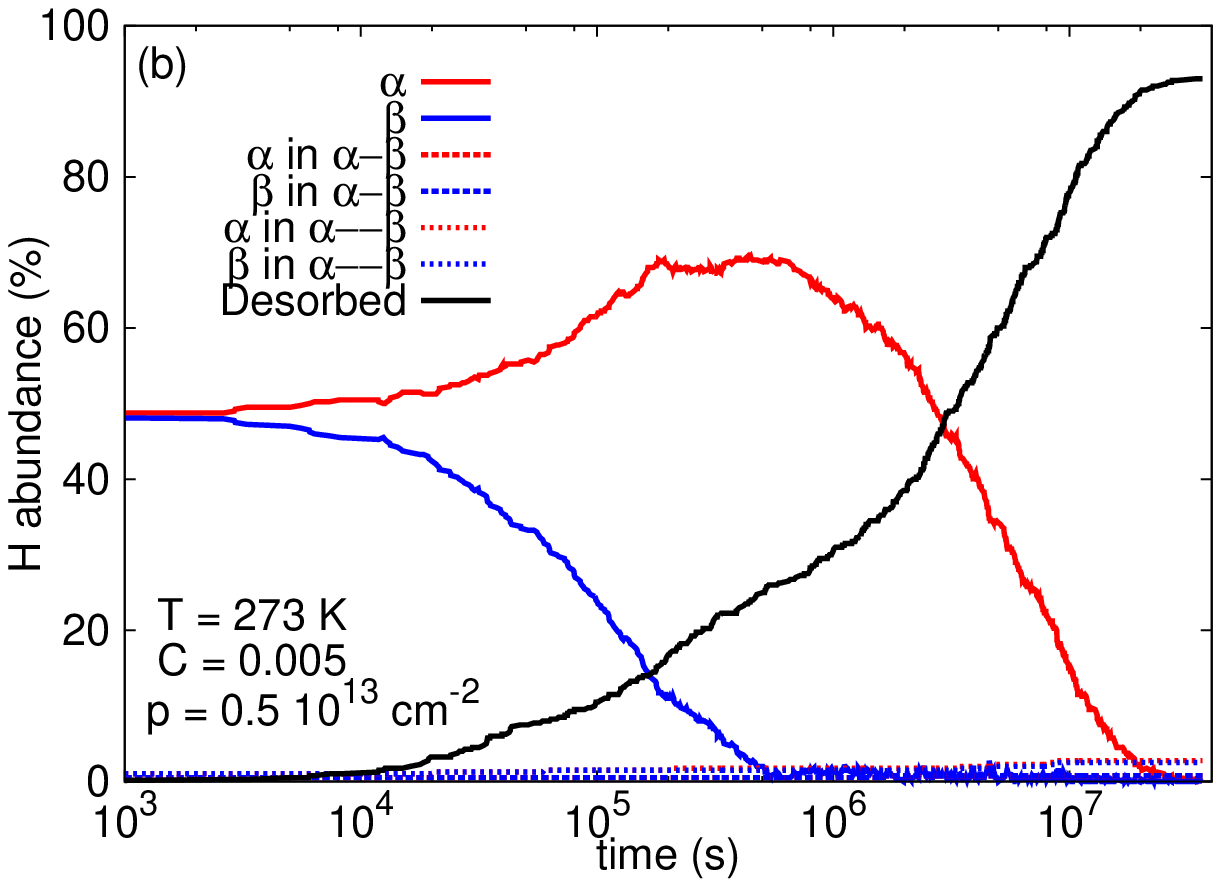}
    \caption{(color online). Same as Fig. 
    \ref{bilayer-abundance}, but for a hole-doped bilayer. The carrier concentration $p$ has been chosen as to make equal the desorption and migration barriers. The H concentration is $C=0.005$ and the temperature is (a) 300 K and (b) 273 K.}
     \label{bilayer-doped1}
\end{figure}

\subsection{Doped graphene bilayer}

We now examine how the dynamics is affected upon doping or carrier concentration variations. DFT calculations for migration and desorption activation barriers in a doped graphene monolayer have been reported in Ref. \onlinecite{Huang11-1}. There it is shown that  doping affects both migration and desorption activation barriers. Following the results in Ref. \onlinecite{Huang11-1} and assuming that the doping reaches the upper layer if induced by the substrate or by a field-effect configuration, we have considered the case of hole doping where the desorption barrier becomes higher than the migration one (this already happens for very small concentration of carriers\cite{Huang11-1}). The dynamics in the case of electron doping is essentially similar to the one in the neutral case, only changing the time scales and will not be explored here.

In Fig. \ref{bilayer-doped1} we plot the time evolution of the relative H abundance for a hole concentration of $p \approx 0.5\, 10^{13}$ cm$^{-2}$ at room temperature (a)  and $T=273$ K (b). According to Ref. \onlinecite{Huang11-1},  this small doping changes the activation barriers so as to make the desorption and migration barriers alike\cite{Huang11-1} in our case. Similarly to the neutral bilayer, after a certain time (which depends on temperature), all H atoms are hosted by the same sublattice. Now, however, migration starts playing a significant role since the desorption is hindered and the concentration of H atoms on the $\beta$ sublattice increases at the expense of the atoms initially adsorbed on the $\alpha$ sublattice. Importantly, instead of desorbing at long times, some H atoms eventually form dimers or clusters ($\approx 10$\% for the chosen concentration of H).  Incidentally, as discussed in Sec. \ref{bilayers},  had we considered the desorption barriers to be $|E_{\rm c}^{\alpha(\beta)}|$, the dynamics of H on a neutral bilayer would have been qualitatively similar to the one shown in Fig. \ref{bilayer-doped1}.

\begin{figure}
    \centering
        \includegraphics[width=3.3in]{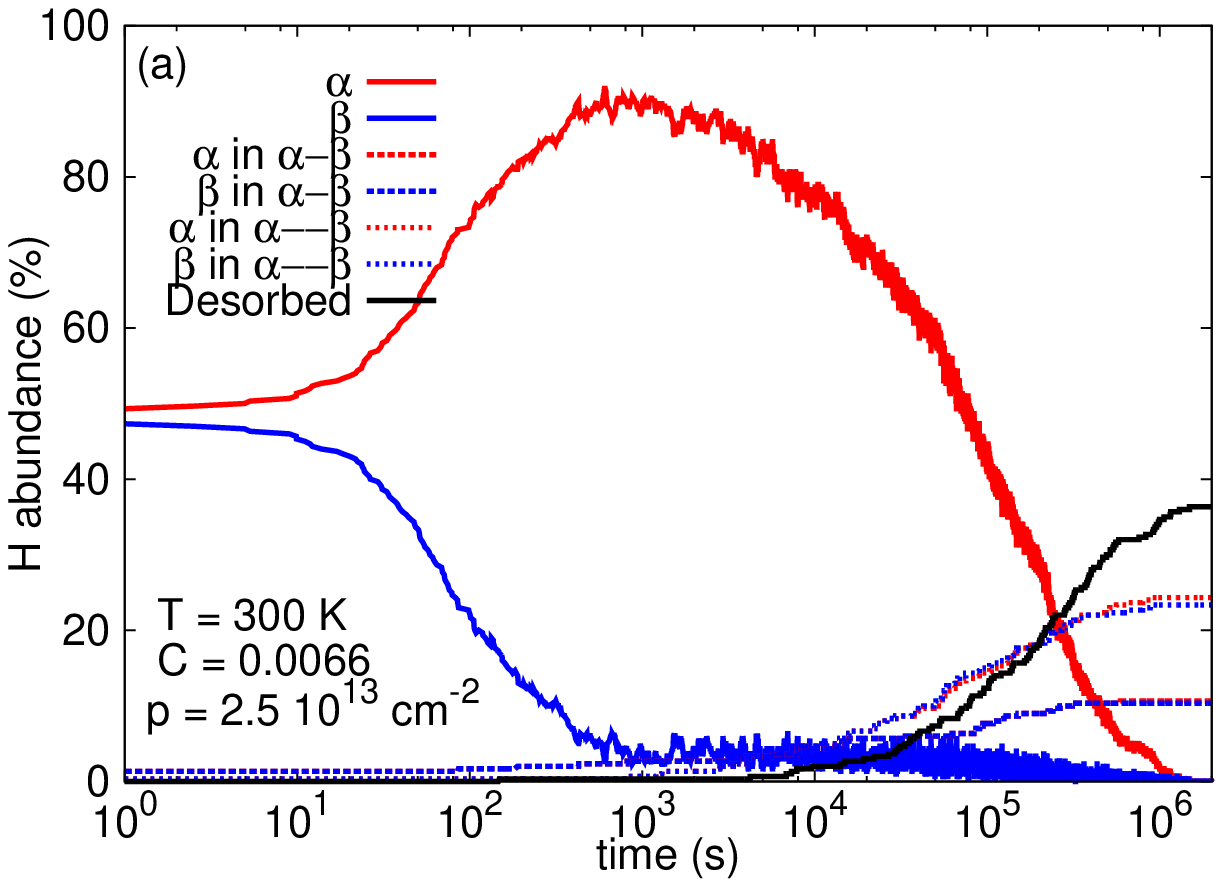}
       \includegraphics[width=3.3in]{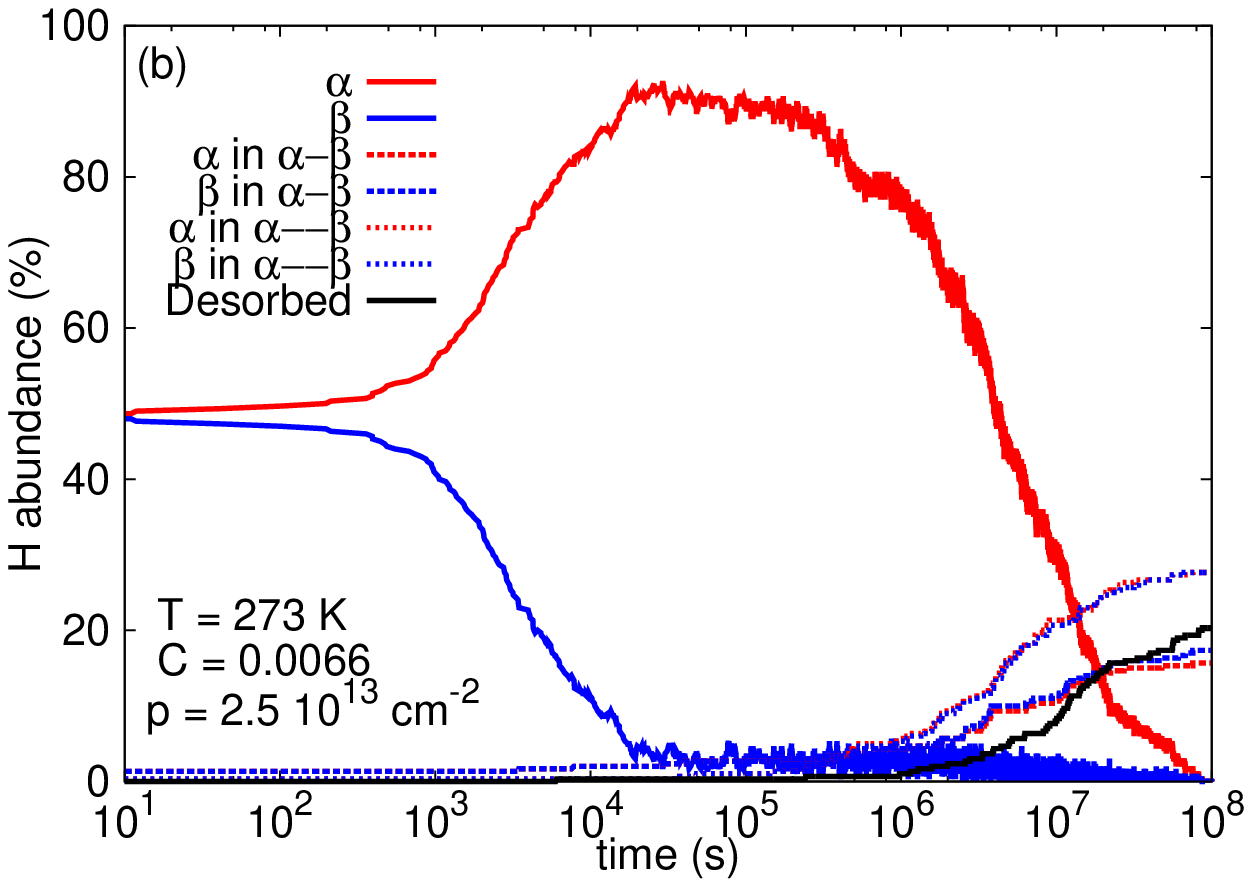}
    \caption{(color online). Same as Fig. 
    \ref{bilayer-doped1}, but for a larger carrier concentration $p$  which  increases even further the desorption barriers and decreases the migration barriers. Two different temperatures are shown: (a) $T=300$ K and (b) $T=273$ K. The concentration of H has been here set to $C=0.0066$.}
     \label{bilayer-doped2}
\end{figure}

In Fig. \ref{bilayer-doped2} we show the time evolution of the H abundance for a larger carrier concentration of  $p \approx 2.5\, 10^{13}$ cm$^{-2}$, again at room temperature (a) and at a lower temperature (b).  A smaller percentage of H atoms are lost now, the rest of them remaining adsorbed on the surface in the form of dimers or clusters  at long times. On changing the initial concentration of H atoms (not shown) the ratio of desorbed to adsorbed atoms logically changes (the larger the concentration, the smaller the percentage of desorbed atoms). Interestingly, even at room temperature, there is now a time window from approximately 1 min to several days where a significant fraction of H atoms remain adsorbed on the same sublattice, coexisting with dimers or clusters at longer times. Notice that $\alpha--\beta$ dimers are more abundant than $\alpha-\beta$ ones, which can be understood since their formation probability is larger due to the larger capture radius. As in the previous cases, all time scales increase on lowering the temperature [see Fig. \ref{bilayer-doped2}(b)].

\section{Conclusions and final considerations}

Our combination of DFT calculations and OKMC simulations have shown that a selective sublattice adsorption of atomic H can be realized on the surface of graphene bilayers for low concentrations. Although, over time, H atoms eventually desorb and/or form clusters, depending on the doping conditions of the substrate,  one can always find a time window where the selectivity persist and allows for measurements of the expected ferromagnetic properties of this system. The time window becomes large enough for measurements or even  practical applications upon lowering the temperature just a few degrees down to $0^{\rm o}$C. 

All the calculations presented here have been carried out for bilayer graphene. However,  given the weak interaction between layers, no qualitative changes are expected for multilayer graphene or graphite. Finally, we should stress that, although the accuracy of DFT calculations is questionable down to the meV range, the uncertainty only affects overall times scales which, as shown, can be easily tuned with temperature or doping. 

\section{Acknowledgments}

This work was supported by MINECO under Grants Nos. FIS2013-47328 and FIS2012-37549, by CAM under Grants Nos. S2013/MIT-3007, P2013/MIT-2850, and by Generalitat Valenciana under Grant PROMETEO/2012/011. The authors acknowledge I. Brihuega for many discussions and for sharing unpublished experimental results with us and J. Soler for enlightening discussions. The authors thankfully acknowledge the computer resources, technical expertise, and assistance provided by the Centro de Computaci\'on Cient\' ifica of the Universidad Aut\'onoma de Madrid and by the Supercomputing and Visualization Center of Madrid (CeSViMa).




\end{document}